\documentstyle[preprint,aps]{revtex}
\begin{document}
 
\draft
 
\title{\large \bf Band Husimi Distributions and the\\ Classical-Quantum 
Correspondence on the Torus}
 
\author{\bf Itzhack Dana\\}
\address{
Department of Physics, Bar-Ilan University, Ramat-Gan 52900, Israel\\} 
\author{\bf Yaakov Rutman and Mario Feingold\\}
\address{
Department of Physics, Ben-Gurion University, Beer-Sheva 84105, Israel\\}
\maketitle

\newpage
 
\begin{abstract}
 
Band Husimi distributions (BHDs) are introduced in the quantum-chaos
problem on a toral phase space. In the framework of this phase space, a
quantum state must satisfy Bloch boundary conditions (BCs) on a torus and
the spectrum consists of a finite number of levels for given BCs. As the
BCs are varied, a level broadens into a band. The BHD for a band is
defined as the uniform average of the Husimi distributions for all the
eigenstates in the band. The generalized BHD for a set of adjacent bands
is the average of the BHDs associated with these bands. BHDs are shown to
be closer, in several aspects, to classical distributions than Husimi
distributions for individual eigenstates. The generalized BHD for two
adjacent bands is shown to be approximately conserved in the passage
through a degeneracy between the bands as a nonintegrability parameter is
varied. Finally, it is shown how generalized BHDs can be defined so as to
achieve physical continuity under small variations of the scaled Planck
constant. A generalization of the topological (Chern-index)
characterization of the classical-quantum correspondence is then obtained.\\ 

\end{abstract}
 
\pacs{PACS numbers: 05.45.+b, 03.65.Sq}

\begin{center}{\bf I. INTRODUCTION}\\ \end{center}

The main objective of ``quantum chaos" is to understand the correspondence
between classically nonintegrable systems and their quantum counterparts
in the semiclassical limit \cite{qc}. During the last two decades,
significant progress has been made in the study of this correspondence
with the discovery of phenomena such as dynamical localization
\cite{qc,fish,she}, ``scarring" of eigenstates by unstable periodic orbits
\cite{qc,hel,bog,ber}, and statistical properties of the eigenspectrum
\cite{qc,sta}. However, the relation between classical phase-space
structures and corresponding quantum-dynamical entities is still far from
being completely understood.\\

In this paper, we introduce quantum-mechanical distributions which, in the
semiclassical limit, are expected to approach in a natural way classical
distributions on both regular and chaotic phase-space structures. This
will be done in the framework of a toral phase space, where, as shown in a
recent series of works \cite{le1,le2,da1}, some interesting new insights
in the quantum-chaos problem can be achieved. Quantum dynamics can be
reduced to a torus if two conditions are satisfied (see Sec. II for more
details). First, the classical map for the system is strictly periodic in
all the phase-space coordinates. The simplest nonintegrable system
possessing this property is the ``kicked-Harper" (KH) model
\cite{le1,le2,da1,tkh,lem,dai,da2,kh} with Hamiltonian: 
\begin{equation}\label{kh} 
H = A\cos (v)+A\cos (u)\sum_{s=-\infty}^{\infty}\delta (t/\tau -s)\ ,
\end{equation} 
where $u$ and $v$ are dimensionless conjugate phase-space variables (with
Poisson bracket $\{ u,\ v\} =1/I$, $I$ being some classical action), $A$
is the amplitude, and $\tau$ is the time period. The system (\ref{kh}) is
exactly related \cite{da2} to the problem of periodically kicked charges
in a uniform magnetic field under resonance conditions \cite{zas}. In the
limit $\tau\rightarrow 0$, (\ref{kh}) reduces to the integrable Harper
Hamiltonian $H_0=A\cos (u)+A\cos (v)$. The transition from integrable to
chaotic phase-space structure, as the dimensionless classical parameter
$\gamma =A\tau /(2\pi I)$ is increased from 0, is shown in Fig. 1. The
second condition for quantum dynamics on a torus is that a scaled $\hbar$,
denoted here by $\rho$ ($[{\hat u},\ {\hat v}]=2\pi i\rho =i\hbar /I$),
assumes rational values, $\rho =q/p$ ($q$ and $p$ are coprime integers).
The admissible quantum states are then those which satisfy Bloch
quasiperiodic boundary conditions (BCs) on the torus, see Sec. II. The
energy or quasienergy spectrum consists of precisely $p$ levels, and, as
the BCs are varied, each of these levels spans a band.\\
 
The advantage of this framework is that it allows for a characterization
of the classical-quantum correspondence by means of integer topological
invariants, the Chern indices \cite{le1,le2,da1}, associated with the $p$
bands. The Chern index $\sigma$ for a band is analogous to the quantum
Hall conductance carried by a magnetic band in a perfect crystal
\cite{qh,avr,sim,da3,daz,wil1,wil2,wil3}, and is a measure of the
sensitivity of the eigenstates in the band to variations in the BCs
\cite{le1,le2,da1,a}. For $q=1$, the toral phase space coincides with the
basic unit cell of periodicity of the system. In this case, where the
classical-quantum correspondence can be established in the simplest and
most natural way, $\sigma$ can assume, in principle, all values. Several
arguments \cite{le1,le2,a}, supported by numerical evidence, then indicate
that if the Husimi distribution of an eigenstate is localized, in a
semiclassical regime ($\rho\ll 1$), on classical regular orbits [e.g.,
Kol'mogorov-Arnol'd-Moser (KAM) tori or periodic orbits] the corresponding
band has $\sigma =0$. On the other hand, eigenstates whose Husimi
distribution is spread over the classical chaotic region should correspond
to bands with $\sigma\neq 0$. The transition from a nearly-integrable
regime [e.g., Fig. 1(a)], where almost all $\sigma =0$, to a fully chaotic
regime [e.g., Fig. 1(d)], where almost all $\sigma\neq 0$, as a
nonintegrability parameter is varied, takes place via degeneracies between
adjacent bands. In the passage through a degeneracy between bands $b$ and
$b'$, the Chern indices $\sigma_b$ and $\sigma_{b'}$ change, respectively,
by $\pm\Delta\sigma$, where, generically \cite{sim,berr},
$\vert\Delta\sigma\vert =1$. A ``diffusion" of Chern indices \cite{wil}
occurs then in the transition above.\\

Despite this characterization of the classical-quantum correspondence for
small $\rho$, the eigenstates may not be considered close to classical
phase-space structures, strictly speaking, for any $\rho$. This is because
of the following reasons: (a) While the BCs satisfied by a quantum state
have a well-defined physical meaning (see Sec. III), they are of a purely
quantum nature. In particular, the strong dependence of the eigenstates on
the BCs for $\sigma\neq 0$, and, in several cases (see Sec. III), also for
$\sigma =0$, has no classical counterpart. (b) The Husimi distribution of
an eigenstate always assumes $p$ zeros in the torus, see Sec. II. Because
of this fact, an eigenstate cannot tend, in the semiclassical limit, to
the microcanonical uniform distribution on the chaotic region
\cite{le1,le3}. (c) In the general case of $q\neq 1$, the exact
eigenstates may be viewed as arising from quantum tunnelling between
degenerate classical orbits located in $q$ adjacent unit cells
\cite{le2,da1,tkh}. As a consequence, $\sigma$ is always nonvanishing
\cite{da1} and the topological characterization of the classical-quantum
correspondence cannot be extended straightforwardly to this general
case.\\

We show in this paper that a natural way to overcome these difficulties is
simply to average uniformly the Husimi distributions of the eigenstates in
a band over all the BCs. In what follows, we refer to the result of this
averaging as the {\it band Husimi distribution} (BHD). The BHDs turn out
to be closer, in several aspects, to classical distributions than Husimi
distributions for individual eigenstates. It is well known \cite{bog,ber}
that smoothing a probability distribution over a range $\Delta E$ of
energy levels has the effect of washing out purely quantum structures such
as scars. As $\Delta E$ is increased, this effect increases and the
smoothed probability distribution (e.g., the spectral Wigner function in
Ref. \cite{ber}) becomes closer to a classical distribution. In our case
the smoothing is done not over a range of {\it discrete} energy levels but
over the {\it continuous} range of one band, corresponding essentially to
a {\it single} level in the framework of a toral phase space. This is thus
the {\it minimal} smoothing in this framework, and it is performed just
for the sake of eliminating the purely quantum effects of individual BCs.
The BHD may be viewed as the representative Husimi distribution for a
level in the torus. In several important cases, we shall find it necessary
to generalize the concept of BHD by considering the average of BHDs
associated with a set of adjacent bands.\\

This paper is organized as follows. In Sec. II, we summarize the relevant
known facts about quantum dynamics on a torus \cite{le1,le2,da1}. In Sec.
III, the concept of BHD for a single band is introduced. In Sec. IV, the
concepts of generalized band (set of adjacent bands) and the associated
BHD are introduced and studied. In Sec. V, the generalized BHD for two
adjacent bands near a degeneracy of these bands is studied. In Sec. VI, we
show how to define generalized BHDs for $q\neq 1$ in order to achieve
physical continuity under small variations in $\rho$. In this way, the
topological characterization of the classical-quantum correspondence is
extended to the general case of $q\neq 1$. Conclusions are presented in
Sec. VII.\\

\begin{center}{\bf II. QUANTUM DYNAMICS ON A TORUS}\\ \end{center}

We provide here some relevant background on quantum dynamics on the torus
\cite{le1,le2,da1} using, for later convenience, the notation of Ref.
\cite{da1}, where a general formulation of the problem was presented. 
Consider a classical area-preserving map strictly periodic in the phase
space $(u,\ v)$ with a $2\pi \times 2\pi$ unit cell, which is the basic
torus $T^2$. The one-step quantum evolution operator corresponding to the
classical map is ${\hat U}({\hat u},\ {\hat v})$, where $[{\hat u},\ {\hat
v}]=2\pi i\rho$. If $\rho$ is a rational number, $\rho =q/p$ ($q$ and $p$
are coprime integers), there exists a pair of ``smallest" commuting
phase-space translations
\begin{equation}\label{ptc} 
{\hat D}_1=e^{i{\hat u}/\rho}, \ \ \ \ \ {\hat D}_2=e^{ip{\hat v}}\ .
\end{equation} 
Because of $[{\hat u},\ {\hat v}]=2\pi i\rho$, ${\hat D}_1$ (${\hat D}_2$)
is a translation by $2\pi$ ($2\pi q$) in the $v$ ($u$) direction. Since
${\hat U}({\hat u},\ {\hat v})$ is periodic in both ${\hat u}$ and ${\hat
v}$ with period $2\pi$, it commutes with ${\hat D}_1$ and ${\hat D}_2$.
There exist therefore simultaneous eigenstates $\Psi$ of ${\hat U}$,
${\hat D}_1$, and ${\hat D}_2$: 
\[ 
{\hat U} \vert\Psi_{b,{\bf w}}\rangle = e^{-i\omega_b({\bf w})}
\vert\Psi_{b,{\bf w}}\rangle\ ,
\] 
\begin{equation}\label{esD12} 
{\hat D}_1\vert\Psi_{b,{\bf w}}\rangle =e^{iw_1/\rho}\vert\Psi_{b,{\bf
w}} \rangle \ ,\ \ \ \ \ {\hat D}_2\vert\Psi_{b,{\bf w}}\rangle =
e^{ipw_2} \vert\Psi_{b,{\bf w}}\rangle \ ,
\end{equation} 
where $b$ is a ``band" index (see below), $\omega_b({\bf w})$ is the
quasienergy, and ${\bf w}=(w_1,\ w_2)$ is a Bloch wavevector varying in
the ``Brillouin zone" (BZ) $0\leq w_1<2\pi \rho$, $0\leq w_2<2\pi /p$. 
Now, for each given value of ${\bf w}$, Eqs. (\ref{esD12}) can be
interpreted as quasiperiodic {\it boundary conditions} (BCs) satisfied by
the eigenstates $\vert\Psi_{b,{\bf w}}\rangle$ in the ``quantum" toral
phase space $T^2_Q$: $0\leq u<2\pi q$, $0\leq v<2\pi$. It can be shown
\cite{da1} that, for each ${\bf w}$, the quasienergy spectrum consists
precisely of $p$ levels $\omega_b({\bf w})$, $b=1,...,\ p$. As the BCs are
varied (by varying ${\bf w}$ in the BZ), each level broadens into a
``band".\\

In the absence of band degeneracies [$\omega_b({\bf w})\neq
\omega_{b'}({\bf w})$ for all ${\bf w}$ and $b'\neq b$],
$\vert\Psi_{b,{\bf w}}\rangle$ must be periodic in the BZ up to a constant
phase factor depending, in general, on ${\bf w}$. The phase of
$\vert\Psi_{b,{\bf w}}\rangle$ may be chosen so that $\vert\Psi_{b,{\bf
w}}\rangle$ will be exactly periodic in one direction, say $w_1$, but then
it will be periodic in $w_2$ only up to a phase factor which, in its
simplest form, has a phase that is linear in $w_1$ \cite{da1}:  
\begin{eqnarray}
\vert\Psi_{b,w_1+2\pi \rho ,w_2}\rangle & = & \vert\Psi_{b,{\bf
w}}\rangle \ , \label{per1}\\ \vert\Psi_{b, w_1,w_2+2\pi /p}\rangle &
= & \exp (i\sigma_bw_1/\rho)\vert\Psi_{b,{\bf w}}\rangle\
. \label{per2}
\end{eqnarray} 
Here the constant $\sigma_b$ must be an integer in order for Eq.
(\ref{per2}) to be consistent with Eq. (\ref{per1}). It is easy to see
from Eqs. (\ref{per1}) and (\ref{per2}) that $2\pi\sigma_b$ is the total
phase change of $\vert\Psi_{b,{\bf w}}\rangle$ when going around the BZ
boundary counterclockwise. This phase change is independent on phase
transformations of $\vert\Psi_{b,{\bf w}}\rangle$. In fact, the integer
$\sigma_b$ is a topological number, the {\it Chern index}, which can be
expressed in a form manifestly invariant under phase transformations [see
Ref. \cite{le1} and expression (\ref{chb}) below].\\

The general form of $\vert\Psi_{b,{\bf w}}\rangle$ in the
$v$-representation is \cite{da1}
\begin{equation}\label{QES}
\Psi_{b,{\bf w}}(v) = \langle v\vert\Psi_{b,{\bf w}}\rangle =
\sum_{m=0}^{p-1}\phi_b(m;\ {\bf w}) \psi_{w_1+2\pi m\rho ,w_2}(v)\ ,
\end{equation} 
where $\phi_b(m;\ {\bf w})$ are expansion coefficients and $\psi_{\bf
w}(v)$ are $kq$-functions \cite{zak},
\begin{equation}\label{kq} 
\psi_{\bf w}(v) = \sum_{l=-\infty}^{\infty}\exp (ilw_1/q)\delta
(v-w_2+2\pi l/p)\ .
\end{equation}
Without the label $b$, (\ref{QES}) is the most general expression of a
quantum state in the $v$-representation for given BCs (i.e., given ${\bf
w}$). A completely analogous expression is obtained in the conjugate ($u$)
representation by Fourier-transforming (\ref{QES}). These expressions
imply, essentially, a discretization of the toral phase space $T^2_Q$ into
$p^2$ points, $u_{m_1}=w_1+2\pi m_1\rho$, $v_{m_2}=w_2+2\pi m_2/p$, $m_1,\
m_2=0,...,\ p-1$, with $\phi (m_1;\ {\bf w})$ being the probability
amplitude for $u$ to assume the value $u_{m_1}$. The corresponding
amplitude for $v=v_{m_2}$ is the discrete Fourier transform of $\phi
(m_1;\ {\bf w})$. This discretization was first studied by Hannay and
Berry \cite{hb} in the special case of $q=1$ ($T^2_Q=T^2$) and ${\bf w=0}$
(strict periodicity). They used the $p$ amplitudes $\phi (m;\ {\bf w=0})$
to investigate the evolution of a quantum state in the cat maps.\\

A most useful representation of $\vert\Psi_{b,{\bf w}}\rangle$ is the
coherent-state representation $\Psi_{b,{\bf w}}({\bf R})=\langle {\bf
R}\vert\Psi_{b,{\bf w}}\rangle$, where ${\bf R}\equiv (u,\ v)$ and
$\vert {\bf R}_0\rangle$ is a coherent state,
\begin{equation}\label{cs}
\langle v\vert {\bf R}_0\rangle = \left (\frac{\alpha ^2}{2\pi
^2\rho}\right ) ^{1/4} \exp\left [ -\frac{\alpha ^2(v-v_0)^2}{4\pi
\rho}- \frac{i u_0}{2\pi \rho}(v-v_0/2)\right ]\ .
\end{equation}
Here $\alpha$ is the ``squeezing parameter'', related to the parameters of
the harmonic oscillator for which the quantum evolution of the state
(\ref{cs}) is nondispersive. An equivalent expression for a coherent state
depends on $u_0$ and $v_0$ only through the complex number $z_0=u_0/\alpha
-i\alpha v_0$ (see, e.g., Refs. \cite{le3,kmw}), giving an analytic
representation $\Psi_{b,{\bf w}}(z)$. For future convenience, however, we
shall use the representation $\Psi_{b,{\bf w}}({\bf R})$, choosing, as in
Ref. \cite{le3}, the symmetric value $\alpha =1$.\\

Important properties of $\Psi_{b,{\bf w}}({\bf R})$ concern its zeros, in
both the ${\bf R}$ and ${\bf w}$ variables. At fixed ${\bf w}$,
$\Psi_{b,{\bf w}}({\bf R})$ always assumes exactly $p$ zeros ${\bf R}={\bf
R}_{0,j}({\bf w})$ ($j=1,...,\ p$) in $T^2_Q$ (counting possible but
nongeneric zero multiplicities). This property was proven in Ref. 
\cite{le3} for $q=1$ and $\alpha =1$, but it can be easily generalized to
all $q$ and $\alpha$, using Eq. (\ref{QES}), together with Eqs. (\ref{kq}) 
and (\ref{cs}). It is also well known \cite{le3} that the $p$ zeros ${\bf
R}_{0,j}({\bf w})$, like the $p$ amplitudes $\phi_b(m;\ {\bf w})$ in
(\ref{QES}), completely determine the wavefunction $\Psi_{b,{\bf w}}({\bf
R})$. At fixed ${\bf R}$, the number $N_0({\bf R})$ of zeros ${\bf w}={\bf
w}_0({\bf R})$ of $\Psi_{b,{\bf w}}({\bf R})$ in the BZ is not smaller
than $\vert\sigma_b\vert$ \cite{note}. These properties imply that the
$p$ zeros ${\bf R}_{0,j}({\bf w})$ must wind around $T^2_Q$ at least
$\vert\sigma_b\vert$ times when ${\bf w}$ is varied in the BZ. Thus, if
for some ${\bf R}={\bf R'}$ $\Psi_{b,{\bf w}}({\bf R'})$ does not vanish
in the BZ, $\sigma_b=0$. We shall refer to the union of all ${\bf R'}$ as
the {\it localization domain} of $\Psi_{b,{\bf w}}({\bf R})$, for reasons
that will become clear in the next paragraph. As ${\bf w}$ is varied over
the entire BZ, the $p$ zeros ${\bf R}_{0,j}({\bf w})$ never enter this
domain (see also Ref. \cite{a}).\\

In a nearly-integrable situation [e.g., very small $A$ in the KH model
(\ref{kh})] and in a semiclassical regime ($\rho\ll 1$), most eigenstates
$\Psi_{b,{\bf w}}({\bf R})$ will be localized on regular classical orbits
and the corresponding bands are usually very narrow (almost independent of
${\bf w}$). A good approximation to the Husimi probability distribution
$\vert\Psi_{b,{\bf w}}({\bf R})\vert ^2$ in its localization region is
given by \cite{le1,kur}
\begin{equation}\label{res}
\vert\Psi_{b,{\bf w}}({\bf R})\vert ^2 \approx \frac{{\cal N}}{V({\bf
R})}\exp \left\{ -\frac{1}{2\pi\rho} \left [ \frac{{\bar H}_0({\bf
R})-E_b}{V({\bf R})}\right ]^2 \right\} \ ,
\end{equation}
where ${\bar H}_0({\bf R})$ is the energy function for an effective
integrable Hamiltonian $H_0$ in this nearly-integrable situation, $E_b$ is
an energy eigenvalue of $H_0$ approximating the very narrow band
[$E_b\approx \omega_b({\bf w})\hbar /\tau$], $V({\bf R})$ is the
phase-space velocity for $H_0$, and ${\cal N}$ is a normalization
constant. Rel. (\ref{res}) manifestly shows that the eigenstates are quite
insensitive to variations in ${\bf w}$ in the region of phase space where
they are localized. In particular, the $p$ zeros ${\bf R}_{0,j}({\bf w})$
should never enter this region, thus implying a finite localization domain
(as defined above). We therefore expect that eigenstates that are
localized on regular classical orbits for $\rho\ll 1$ should belong to
bands with $\sigma_b=0$ (see also Sec. III).\\

It is important to remark here that the value $\sigma_b=0$ may occur only
for $q=1$, since $\sigma_b$ has to satisfy the Diophantine equation
$p\sigma_b+q\mu_b=1$, where $\mu_b$ is a second integer \cite{da1,da3}.
For $q\neq 1$, the exact eigenstates are quite ``nonclassical", since they
may be viewed as arising from quantum tunnelling between $q$ degenerate
classical orbits located in the $q$ adjacent unit cells defining $T^2_Q$
\cite{le2,da1,tkh}. Thus, even in a completely integrable situation, the
$p$ zeros ${\bf R}_{0,j}({\bf w})$ always cover the torus $T^2_Q$ when
${\bf w}$ varies in the BZ \cite{le2}.\\

In a strongly chaotic situation, where a typical orbit fills the entire
phase space $T^2$, the $p$ zeros ${\bf R}_{0,j}({\bf w})$ for $q=1$ are
expected to be distributed almost uniformly in $T^2_Q=T^2$ \cite{le3} and
to explore the entire $T^2$ as ${\bf w}$ is varied in the BZ \cite{le1}.
Almost all the bands should therefore be characterized by nonvanishing
$\sigma_b$. The transition from a nearly-integrable situation (almost all
$\sigma_b=0$) to a strongly chaotic one (almost all $\sigma_b\neq 0$), as
a nonintegrability parameter is varied, takes place via successive
degeneracies between adjacent bands \cite{le1,le2} (see also the
Introduction).\\

\begin{center}{\bf III. BAND HUSIMI DISTRIBUTIONS}\\ \end{center}

We start this section by considering in more detail the nature of the BCs
(\ref{esD12}) using the KH system (\ref{kh}) as a model. The
nearly-integrable regime for this system corresponds to very small values
of the classical parameter $\gamma =A\tau /(2\pi I)$ [see Fig. 1(a)]. In
this regime, the system is well described by the Harper Hamiltonian
$H_0=A\cos (u)+A\cos (v)$. For $\rho =1/p$, $p$ odd, the energy spectrum
of $H_0$ consists of $p$ bands $E_b({\bf w})$, $b=1,...,\ p$, in order of
increasing energy. Only the central band [$b=(p+1)/2$] has a nonvanishing
Chern index $\sigma_b=1$ \cite{qh}. In the semiclassical regime, $p\gg 1$,
the eigenstates of this band are concentrated on the separatrix orbit [see
Fig. 1(a)], which is not contractible to a point. On the other hand, the
eigenstates of the other bands, that have vanishing Chern indices, are
concentrated on orbits which are contractible to a point.\\

This state of matters persists also for $\gamma$ not very small, when the
separatrix orbit breaks into a stochastic layer, and island chains emerge
from the $H_0$ contractible orbits. For example, for $\rho =1/11$, the
Chern indices $\sigma_b$ appear to be the same as in the Harper case in
the entire interval $\gamma\leq 0.26$ \cite{le2,not}! Namely, only the
central quasienergy band features a nonvanishing Chern index $\sigma_b=1$.
We have studied numerically the sensitivity of the eigenstates to
variations in the BCs (i.e., variations in ${\bf w}$) for several values
of $\gamma$ in the interval above. We find that eigenstates in
$\sigma_b=0$ bands sufficiently ``far" from the central band (i.e., $b$
close to 1 or 11) are indeed almost insensitive to variations in ${\bf w}$
(large localization domain). This is not the case, however, for bands
sufficiently close to the central band. While the Chern index for these
bands vanishes, the sensitivity of the eigenstates to variations in ${\bf
w}$ is quite strong (very small or empty localization domain), almost as
that of eigenstates in the central band. This is clearly illustrated in
Figs. 2 and 3. In general, we expect strong sensitivity to variations in
${\bf w}$ in $\sigma_b=0$ bands that are sufficiently close to
$\sigma_b\neq 0$ bands.\\

This sensitivity has no classical analogue, since ${\bf w}$ is a purely
quantum characterization of the eigenstates, which are concentrated on
different regions of $T^2_Q$ for different values of ${\bf w}$. In fact,
by comparing Eqs. (\ref{esD12}) with the definitions (\ref{ptc}) of the
operators ${\hat D}_1$ and ${\hat D}_2$, we see immediately that ${\bf w}$
is just a {\it quasicoordinate} of ${\bf R}=(u,\ v)$:  
\[ 
w_1\Longleftrightarrow\rho \left ( \frac{u}{\rho}\bmod 2\pi\right )\ ,
\ \ \ \ \ \ w_2\Longleftrightarrow\frac{1}{p}(pv\bmod 2\pi )\ .
\]
To understand this better, consider, for given ${\bf w}={\bf w'}$, the
``displaced" quantum state
\begin{equation}\label{des} 
{\hat D}(z_0)\vert\Psi_{b,{\bf w'}}\rangle =
\exp(z_0{\hat a}^{\dagger}-z_0^{\ast}{\hat a})\ 
\vert\Psi_{b,{\bf w'}}\rangle \ , 
\end{equation} 
where $z_0=u_0+iv_0$ and ${\hat a}=({\hat u}+i{\hat v})/(4\pi\rho )$ is
the annihilation operator. It is easy to verify, using the well known
commutation relation for the phase-space translations ${\hat D}(z_0)$
\cite{per}, that the state (\ref{des}) satisfies the BCs (\ref{esD12})
with the displaced value of ${\bf w}={\bf w'}+{\bf R}_0$ [${\bf
R}_0=(u_0,\ v_0)$]. On the other hand, using the expression (\ref{cs})
(for $\alpha =1$), we find that the coherent-state representation of the
state (\ref{des}) is given by
\begin{equation}\label{hdes} 
\langle {\bf R}\vert {\hat D}(z_0)\vert\Psi_{b,{\bf w'}}\rangle = \exp
[i(uv_0-u_0v)/(4\pi\rho )]\ \Psi_{b,{\bf w'}}({\bf R}-{\bf R}_0)\ .
\end{equation} 
Thus, the shifted function $\Psi_{b,{\bf w'}}({\bf R}-{\bf R}_0)$ is, up
to a phase factor, the coherent-state representation of a quantum state
characterized by the displaced value of ${\bf w}={\bf w'}+{\bf R}_0$. This
is a vivid illustration of the notion of ``quasicoordinate of ${\bf R}$"
for ${\bf w}$.\\

From Eq. (\ref{hdes}) one may get the impression that there is always
strong sensitivity to variations in the BCs. However, since ${\hat
D}(z_0)$ does not commute, for general $z_0$, with the evolution operator
${\hat U}$, the displaced state (\ref{des}) will not be, in general, an
eigenstate. Still, one has the expansion
\[
\langle {\bf R}\vert {\hat D}(z_0)\vert\Psi_{b,{\bf w'}}\rangle = 
\sum_{b'=1}^p c_{b,b'}({\bf R}_0,\ {\bf w})\Psi_{b',{\bf w}}({\bf R}) ,
\]
where the coefficients $c_{b,b'}({\bf R}_0,\ {\bf w})=\langle\Psi_{b',{\bf
w}}\vert {\hat D}(z_0)\vert\Psi_{b,{\bf w'}}\rangle$. Assume, for example,
that $\Psi_{b,{\bf w}}({\bf R})$ is almost insensitive to variations in
${\bf w}$ (large localization domain). Then, from Eq. (\ref{hdes}) and
${\bf w}={\bf w'}+{\bf R}_0$ it follows that the coefficient $c_{b,b}({\bf
R}_0,\ {\bf w})$ should be relatively small for almost all ${\bf w}$ if
$R_0$ is larger than the typical width of the localization domain. If, on
the other hand, $\Psi_{b,{\bf w}}({\bf R})$ is highly sensitive to
variations in ${\bf w}$, the coefficients $c_{b,b'}({\bf R}_0,\ {\bf w})$,
for $b'\approx b$, may be relatively large for ``many" pairs $({\bf R}_0,\
{\bf w})\neq {\bf 0}$. For example, in the cases shown in Figs. 2 and 3,
these pairs include $[{\bf R}_0=(\pi ,\ \pi ),\ {\bf w}=(0,\ \pi /11)]$
[i.e., ${\bf w'}=(\pi /11,\ 0)$; compare Fig. 2(b) with Fig. 2(c) and Fig.
3(b)  with Fig. 3(c)] and $[{\bf R}_0=(\pi ,\ \pi ),\ {\bf w}=(\pi /11,\
\pi /11)]$ [i.e., ${\bf w'}=(0,\ 0)$; compare Fig. 2(a) with Fig. 2(d)].\\

Because of the purely quantum nature of ${\bf w}$ and the BCs, it is
natural to perform a uniform average over ${\bf w}$ in the Brillouin zone
in order to obtain ``more classical" quantities. In this paper, we shall
consider the average of the Husimi probability distribution
$\vert\Psi_{b,{\bf w}}({\bf R})\vert ^2$, giving the {\it band Husimi
distribution} (BHD) for band $b$: 
\begin{equation}\label{BHD}
P_b({\bf R})=\frac{1}{\vert {\rm BZ}\vert} \int_{\rm BZ} d{\bf w}
\vert\Psi_{b,{\bf w}}({\bf R})\vert ^2 \ ,
\end{equation}
where $\vert {\rm BZ}\vert =4\pi ^2q/p^2$ is the area of the Brillouin
zone. The BHD (\ref{BHD}) corresponds to the {\it minimal} smoothing of a
probability distribution in the framework of a toral phase space, namely
the smoothing over the continuous range of a single band. We now show that
this smoothing is sufficient to make the BHD ``more classical" than an
individual Husimi distribution $\vert\Psi_{b,{\bf w}}({\bf R})\vert ^2$ in
several aspects. First, we notice from Eq. (\ref{esD12}) that
$\vert\Psi_{b,{\bf w}}({\bf R})\vert ^2$ is periodic only with unit cell
$T^2_Q$, i.e., the quantum phase space, which differs from the classical
one, $T^2$, whenever $q\neq 1$. Now, using the relation \cite{da1} \[
{\hat D}(-2\pi s)\vert\Psi_{b,{\bf w}}\rangle = \exp (isw_2/\rho
)\vert\Psi_{b,w_1-2\pi s,w_2}\rangle \] ($s$ integer) as well as Eq.
(\ref{hdes}) in (\ref{BHD}), we easily find that
\begin{equation}\label{BHDq} 
P_b({\bf R})=\frac{1}{q}\sum_{s=0}^{q-1}P_b^{(q)}(u+2\pi s,\ v)\ ,
\end{equation}
where $P_b^{(q)}({\bf R})$ is defined as in Eq. (\ref{BHD}), but the
integral is performed over $1/q$ of the BZ, i.e., $0\leq w_1,\ w_2<2\pi
/p$, and $\vert {\rm BZ}\vert$ is replaced by $\vert {\rm BZ}\vert /q$. It
is then clear from Eq. (\ref{BHDq}) that $P_b({\bf R})$, unlike
$\vert\Psi_{b,{\bf w}}({\bf R})\vert ^2$, is periodic with unit cell $T^2$
for {\it general} $q$. This allows one to impose on $P_b({\bf R})$ the
normalization condition
\begin{equation}\label{nr} 
\int_{T^2} d{\bf R} P_b({\bf R}) = 1\ .  
\end{equation} 
This makes $P_b({\bf R})$ analogous to a classical probability
distribution on the classical phase space $T^2$.\\

Second, $\vert\Psi_{b,{\bf w}}({\bf R})\vert ^2$ always assumes $p$ zeros
${\bf R}_{0,j}({\bf w})$ ($j=1,...,\ p$) in $T^2_Q$ (see Sec. II). These
zeros give $\vert\Psi_{b,{\bf w}}({\bf R})\vert ^2$ a rather
``nonclassical" appearance, for example, they do not allow
$\vert\Psi_{b,{\bf w}}({\bf R})\vert ^2$ to approach, in the semiclassical
limit, the microcanonical uniform distribution in the chaotic region
(strong-chaos regime) \cite{le1,le3}. On the other hand, the BHD {\it
never vanishes} in the phase space [$P_b({\bf R})>0$ for all ${\bf R}$ in
$T^2$], simply because the $p$ zeros ${\bf R}_{0,j}({\bf w})$ ($j=1,...,\
p$) generally vary with ${\bf w}$, and, by definition [see Eq.
(\ref{BHD})], a BHD involves an integration over all ${\bf w}$. It is
therefore possible for a BHD to resemble a classical probability
distribution. For example, in Fig. 4 we show the BHD for the case
considered in Fig. 2 ($b=6$). The relatively high probability density near
the hyperbolic points ${\bf R}=(\pi,\ 0),\ (0,\ \pi )$ can be easily
understood from classical considerations. The band $b=6$ corresponds to a
``broken" separatrix orbit (that is, a homoclinic orbit in the chaotic
layer), and the phase-space velocity on this orbit vanishes as one
approaches the hyperbolic points. Accordingly, the approximate formula
(\ref{res}) suggests that the BHD should assume relatively high values
near these points. Fig. 5 shows that the localization region of the BHD
for band $b=5$ is completely different from that of individual Husimi
distributions (see Fig. 3). Although this band has $\sigma_b=0$, it
exhibits strong sensitivity to variations of the BCs. The representative
or dominant localization region for such a band can be found only by
inspecting its BHD. In the strongly chaotic regime ($\gamma\gg 1$) and in
the semiclassical limit, the BHDs are expected to approach the
microcanonical uniform distribution.\\

Finally, consider the special but important case of bands with
$\sigma_b=0$, which is possible only for $q=1$ (see Sec. II). Here, an
eigenstate $\Psi_{b,{\bf w}}({\bf R})$ can always be written as a
symmetry-adapted sum \cite{da3,wil1}
\begin{equation}\label{sas}
\Psi_{b,{\bf w}}({\bf R})=\sum_{l_1,l_2=-\infty}^{\infty} \exp
[-ip(l_1w_1+l_2w_2)] \langle {\bf R}\vert
{\hat D}_1^{l_1}{\hat D}_2^{l_2}\vert\varphi_b\rangle \ ,
\end{equation}
where ${\hat D}_1$ and ${\hat D}_2$ are the basic phase-space translations
(\ref{ptc}) and $\vert\varphi_b\rangle$ is some square-integrable state,
which is analogous to a Wannier function \cite{da3,wil1,wil3}. Inserting
Eq. (\ref{sas}) into Eq. (\ref{BHD}), we easily obtain the exact
expression
\begin{equation}\label{BHD0} 
P_b({\bf R})=\sum_{l_1,l_2=-\infty}^{\infty} \vert\langle u+2\pi l_1,\
v+2\pi l_2\vert\varphi_b\rangle\vert ^2\ .
\end{equation}
While the Wannier function $\langle {\bf R}\vert\varphi_b\rangle$ is not
invariant under gauge transformations in which the eigenstates are
multiplied by phase factors $\exp [i\theta ({\bf w})]$ \cite{wil3}, the
BHD (\ref{BHD0}) is gauge invariant. In a nearly-integrable situation and
in a semiclassical regime, $\vert\langle {\bf R}\vert\varphi_b\rangle\vert
^2$ may be identified with the ``quasi-mode" of Ref. \cite{fau}, which is
well localized on a classical regular orbit and, in the limit
$\rho\rightarrow 0$, tends point-wise to zero outside this orbit. 
Similarly, the BHD (\ref{BHD0}) tends point-wise to zero outside the
periodic repetition of the orbit on all unit cells $(l_1,\ l_2)$. It is
therefore a periodic version of the quasi-mode, appropriate for a toral
phase space. A good approximation to the BHD should be given by the
right-hand side of Eq. (\ref{res}), which, like (\ref{BHD0}), is
essentially independent of ${\bf w}$ and is periodic with unit cell $T^2$.
The difference $Z_{b,{\bf w}}({\bf R})\equiv\vert\Psi_{b,{\bf w}}({\bf
R})\vert ^2-P_b({\bf R})$ is the sum of the overlaps of the quasi-mode
$\langle u+2\pi l_1,\ v+2\pi l_2\vert \varphi_b \rangle$ [in unit cell
$(l_1,\ l_2)$] with a quasi-mode in a different unit cell. The function
$Z_{b,{\bf w}}({\bf R})$ is thus of a purely quantum nature and it is
entirely responsible to the $p$ zeros of $\Psi_{b,{\bf w}}({\bf R})$ and
to the sensitivity of $\Psi_{b,{\bf w}}({\bf R})$ to variations of ${\bf
w}$ for ${\bf R}$ outside the localization domain. This clarifies the
classical nature of the BHD in this case.\\

\begin{center}{\bf IV. GENERALIZED BANDS AND BHDs}\\ \end{center}
  
In several important situations, some of which will be considered in the
next sections, it is necessary to generalize the concept of BHD by
smoothing over more than a single band, namely over a set of $N$ adjacent
bands $b=b_1,...,\ b_N$. This gives the {\it generalized} BHD
\begin{equation}\label{BHDg} 
P_{b_1-b_N}({\bf R})=\frac{1}{N}\sum_{b=b_1}^{b_N} P_b({\bf R})\ .
\end{equation}
The additional smoothing over bands should give a ``more classical" BHD,
as when smoothing over many levels in a bounded quantum system \cite{ber}. 
The ``maximal" smoothing is, of course, that over all the $p$ bands. From
the completeness of the eigenstates (\ref{QES}), together with the
normalization condition (\ref{nr}), we find in this case that
\[
\frac{1}{p}\sum_{b=1}^p P_b({\bf R}) = \frac{1}{\vert T^2\vert}\ ,
\]
where $\vert T^2\vert =4\pi ^2$ is the area of the classical torus. Thus,
as one could expect, the generalized BHD in this case is just the uniform
distribution over phase space.\\

The set of $N$ adjacent bands can be considered as a single entity, a {\it
generalized band} (GB). One may perform arbitrary linear combinations of
GB eigenstates at fixed ${\bf w}$ to obtain general states satisfying
given BCs (\ref{esD12}) on the torus. While these states are generally not
eigenstates of the evolution operator, they are ``almost stationary" 
provided that the energy or quasienergy width of the GB is sufficiently
small \cite{da1}. The set of all these states, for all ${\bf w}$, is the
space of the GB. A natural starting basis for this space is, of course,
the set of $N$ eigenstates $\Psi_{b,{\bf w}}({\bf R})$, $b=b_1,...,\ b_N$,
at each fixed value of ${\bf w}$. An arbitrary basis will then be given by
\begin{equation}\label{ut}
\overline{\Psi}_{n,{\bf w}}({\bf R}) = \sum_{b=b_1}^{b_N}
B_b^{(n)}({\bf w}) \Psi_{b,{\bf w}}({\bf R})\ ,
\end{equation}
$n=1,...,\ N$. To ensure orthonormality of the basis (\ref{ut}) in the new
``band index" $n$, the coefficients $B_b^{(n)}({\bf w})$ must form a
unitary matrix. Obviously, the states (\ref{ut}) satisfy the BCs
(\ref{esD12}). In addition, it is natural to require that these states
will satisfy quasiperiodicity conditions in ${\bf w}$, analogous to those
of Eqs. (\ref{per1}) and (\ref{per2}), with well-defined Chern indices
${\bar \sigma}_n$. Clearly, this will be the case only if the matrix
$B_b^{(n)}({\bf w})$ in Eq. (\ref{ut}) satisfies these conditions with
Chern indices $\sigma_{b,n}={\bar \sigma}_n-\sigma_b$. The determinant of
this matrix must be strictly periodic in ${\bf w}$, otherwise it will
vanish at some ${\bf w}$ (see Sec. II and note \cite{note}), which cannot
happen since the matrix is unitary. It follows from this that
\begin{equation}\label{st}
\sigma_{\rm GB}\equiv\sum_{b=b_1}^{b_N}\sigma_b = \sum_{n=1}^{N}{\bar
\sigma}_n\ .
\end{equation}
In other words, the total Chern index $\sigma_{\rm GB}$ of the GB is
``conserved" under the basis transformation in Eq. (\ref{ut}).\\

Another ``conservation" law following from Eq. (\ref{ut}) is
\begin{equation}\label{sqe}
\sum_{b=b_1}^{b_N} \vert\Psi_{b,{\bf w}}({\bf R})\vert ^2 =
\sum_{n=1}^{N}\vert\overline{\Psi}_{n,{\bf w}}({\bf R})\vert ^2 \ .
\end{equation}
In particular, Eq. (\ref{sqe}), when integrated over ${\bf w}$, implies
that the generalized BHD (\ref{BHDg}) can be calculated using an arbitrary
basis (\ref{ut}).\\

An important case is when one can find a new basis (\ref{ut}) whose Chern
indices ${\bar \sigma}_n$ all vanish. Before discussing the meaning of
this case, we first determine the conditions that need to be satisfied to
make it possible. Because of Eq. (\ref{st}), a necessary condition is
clearly that $\sigma_{\rm GB}=0$. This condition is also sufficient, since
if $\sigma_{\rm GB}=0$ one can always find a unitary matrix
$B_b^{(n)}({\bf w})$ with Chern indices $\sigma_{b,n}=-\sigma_b$ (implying
that ${\bar \sigma}_n=0$ for all $n$). In fact, one can simply choose
$B_b^{(n)}({\bf w})$, $n=1,...,\ N$, as the $N$ orthonormal eigenvectors
of a $N\times N$ Hermitian matrix that is strictly periodic in ${\bf w}$
and whose $N$ homotopic invariants (Chern indices) are $-\sigma_b$,
$b=b_1,...,\ b_N$. Such a matrix can be explicitly constructed for any
given set of integers $\sigma_b$ with $\sigma_{\rm GB}=0$ \cite{avr}. It
is worthwhile to stress here that due to the general Diophantine relation
$p\sigma_b+q\mu_b=1$ \cite{da1,da3} the condition $\sigma_{\rm GB}=0$ can
be satisfied only if $N$ is a multiple of $q$, i.e., the minimal $N$ is
$N=q$.\\

The fact that one can find a new basis (\ref{ut}) with ${\bar \sigma}_n=0$
for all $n$ means that the GB can be viewed as ``weakly sensitive" to
variations in the BCs, despite the fact that the original Chern indices
$\sigma_b$ may be all different from zero. This can be expressed in a more
precise way using Eq. (\ref{sqe}). Since ${\bar \sigma}_n=0$, one can
assume that $\overline{\Psi}_{n,{\bf w}}({\bf R})$ has a finite
localization domain (see Sec. II). Eq. (\ref{sqe}) then implies that the
function $\sum_{b=b_1}^{b_N} \vert\Psi_{b,{\bf w}}({\bf R})\vert ^2$,
characterizing the GB, has also a finite localization domain. In this
sense, the GB exhibits ``weak" sensitivity to variations in the BCs.
Nonzero values of the Chern indices $\sigma_b$ in this case only mask the
true nature of the GB, which is best described in terms of the new basis.
One can thus say that a GB consisting of $q$ adjacent bands with
$\sigma_{\rm GB}=0$ is analogous to a band with $\sigma_b=0$ in the case
of $q=1$. In Sec. VI, these ideas will be further developed in order to
generalize the Chern-index characterization of the classical-quantum
correspondence to the case of $q\neq 1$.\\

\begin{center}{\bf V. BHDs NEAR DEGENERACIES}\\ \end{center}

In this section, we show that the generalized BHD for two adjacent bands
(to be denoted, for simplicity, by $b=1,\ 2$) is approximately conserved
as a nonintegrability parameter $\gamma$ is slightly varied through a
degeneracy point $\gamma_0$ of these bands, i.e., $\omega_1({\bf
w}_0)=\omega_2({\bf w}_0)$ at $\gamma =\gamma_0$, where ${\bf w}_0$ is
some isolated value of ${\bf w}$. This despite the fact that the separate
BHDs of the two bands usually change significantly under such variation.\\

Formally, the unitary evolution operator can be written as ${\hat U}=\exp
[-i{\hat G}(\gamma )]$, where ${\hat G}(\gamma )$ is a Hermitian operator.
The quasienergy states (\ref{QES}) are eigenstates of ${\hat G}(\gamma )$
with eigenvalues $\omega_b({\bf w})$. Consider two values of $\gamma$,
$\gamma_1$ and $\gamma_2$, very close to $\gamma_0$ and such that
$\gamma_1<\gamma_0<\gamma_2$, and let us denote by $\vert\Psi_{b,{\bf
w}}^{(j)}\rangle$, $j=1,\ 2$, the eigenstates of ${\hat G}(\gamma_j)$. We
assume that for $\gamma$ in the interval $[\gamma_1,\ \gamma_2]$
(containing the degeneracy point), and for all ${\bf w}$, the distance
between $\omega_1({\bf w})$ and $\omega_2({\bf w})$ is significantly
smaller than the distance between any of these quasienergies and
$\omega_b({\bf w}$), $b\neq 1,\ 2$. In this case, one can write, to a good
approximation,
\begin{equation}\label{dut} 
\Psi_{b',{\bf w}}^{(2)}({\bf R})\approx \sum_{b=1}^2 B_b^{(b')}({\bf
w}) \Psi_{b,{\bf w}}^{(1)}({\bf R})\ ,
\end{equation} 
where the expansion coefficients $B_b^{(b')}({\bf w})$, $b,\ b'=1,\ 2$,
form a $2\times 2$ unitary matrix built from the normalized eigenvectors
of the $2\times 2$ Hermitian matrix $G_{b,b'}({\bf w})=\langle\Psi_{b,{\bf
w}}^{(1)}\vert {\hat G}(\gamma_2)\vert\Psi_{b',{\bf w}}^{(1)}\rangle$. Eq.
(\ref{dut}) is thus an approximate special case of Eq. (\ref{ut}).
Nevertheless, the general relation (\ref{st}) holds exactly in this case
as well. It expresses the well known conservation of the total Chern index
$\sigma_1+\sigma_2$ in the passage through a degeneracy point
\cite{sim,berr}.\\

From Eq. (\ref{dut}) we get the relation
\begin{equation}\label{sqw}
\vert \Psi_{1,{\bf w}}^{(1)}({\bf R})\vert ^2 + \vert \Psi_{2,{\bf
w}}^{(1)}({\bf R})\vert ^2\approx \vert \Psi_{1,{\bf w}}^{(2)}({\bf
R})\vert ^2 + \vert \Psi_{2,{\bf w}}^{(2)}({\bf R})\vert ^2\ ,
\end{equation}
which is an approximate special case of Eq. (\ref{sqe}). After integrating
Eq. (\ref{sqw}) over the entire BZ, we obtain the approximate conservation
law for the generalized BHD of the two bands: 
\[ P_{1,2}^{(1)}({\bf R})\approx P_{1,2}^{(2)}({\bf R})\ . \]

As a first, instructive example, we consider the degeneracy between bands
$b=3,\ 4$ in the KH model with $\rho =1/11$ for $\gamma =\gamma_0\approx
0.264$ (see Ref. \cite{le2}). For $\gamma =\gamma_1=0.26<\gamma_0$, the
Chern indices of the two bands are, respectively, $\sigma_3=\sigma_4=0$,
while for $\gamma =\gamma_2=0.2645>\gamma_0$ they change to
$\sigma_3=-\sigma_4=2$ \cite{not}. For $\gamma =\gamma_3\approx 0.2653$,
the Chern indices re-assume the values $\sigma_3=\sigma_4=0$ due to a
second degeneracy between the two bands. Thus, for both $\gamma =\gamma_1$
and $\gamma =\gamma_3$, one can associate Wannier functions $\langle {\bf
R}\vert \varphi_b\rangle$ with these bands, as in Eq. (\ref{sas}). These
functions are expected to be localized on classical regular orbits (tori),
such as those shown in Fig. 1(b). Our numerical results indicate that
$\langle {\bf R}\vert\varphi_b\rangle$ for $b=3$ ($b=4$) at $\gamma
=\gamma_3$ is essentially the same as $\langle {\bf R}\vert\varphi_b
\rangle$ for $b=4$ ($b=3$) at $\gamma =\gamma_1$. This ``exchange" of
Wannier functions in the passage through the degeneracy can be understood
as follows. Sufficiently far from the degeneracy region the bands vary
almost linearly as a function of $\gamma$ (see Fig. 11 in Ref. \cite{le2})
and are well approximated by a primitive semiclassical quantization of the
two tori on which the functions $\langle {\bf R}\vert\varphi_b\rangle$,
$b=3,\ 4$, are localized. Near the degeneracy region, however, the actual
band structure results from an ``avoided crossing" between the two bands
[see Fig. 12(a) in Ref. \cite{le2}], leading to the exchange phenomenon.
In Fig. 6(a), we plot the BHDs for the two bands along the symmetry line
$u=v$ for $\gamma =\gamma_1$. These BHDs exhibit, essentially, the
profiles of the two Wannier functions. A similar plot for $\gamma
=\gamma_2$ (between the two degeneracies) is shown in Fig. 6(b). It is
evident that the BHDs have changed significantly following the small
variation in $\gamma$ from $\gamma_1$ to $\gamma_2$. However, Fig. 6(c)
shows that the generalized BHD for the two bands is conserved to high
accuracy under this variation. This BHD, associated with a GB having a
total Chern index $\sigma_{\rm GB}=0$ (see Sec. IV), corresponds
essentially to a ``superposition" of the two Wannier functions. Between
the two degeneracies, the GB can be still described by Wannier functions,
associated with a new basis (\ref{ut}) with ${\bar \sigma}_n=0$.\\

As a second example, we consider a GB with $\sigma_{\rm GB}\neq 0$ in the
same model ($\rho =1/11$). For $\gamma =\gamma_0\approx 0.3387$, a
degeneracy between the bands $b=5$ and $b=6$ takes place. Due to the
special symmetries of the KH model \cite{le1}, a degeneracy between bands
$b=6$ and $b=7$ must occur at precisely the same value of $\gamma_0$ but
at a different value of ${\bf w}_0$. As $\gamma$ is varied through
$\gamma_0$, the Chern indices of bands $b=5,\ 7$ both change from
$\sigma_b=0$ to $\sigma_b=-1$ while $\sigma_6$ changes from $\sigma_6=1$
to $\sigma_6=3$. The BHDs for all these bands are shown in Fig. 7(a) (for
$\gamma <\gamma_0$) and in Fig. 7(b) (for $\gamma >\gamma_0$). Since the
bands $b=5,\ 6,\ 7$ degenerate at the same value of $\gamma_0$, it is
clear from our previous analysis that only the generalized BHD of all
these three bands can be expected to be approximately conserved in the
passage through the degeneracy point. In fact, Fig. 7(c) shows that this
BHD, associated with a GB having $\sigma_{\rm GB}=1$, is conserved with
significantly better accuracy than in the case of Fig. 6(c).\\

\begin{center}{\bf VI. BHDs UNDER SMALL VARIATIONS OF $\rho$}\\
\end{center}

In this section, we show how to define generalized BHDs for $q\neq 1$ that
are continuous under small variations of $\rho$ on the rationals. The
topological characterization of the classical-quantum correspondence in
the torus is then generalized to $q\neq 1$. We shall use the
renormalization-group approach of Wilkinson \cite{wil2,wil3}, which was
applied to the investigation of the spectrum of a general class of
time-independent Hamiltonians on the torus. In what follows, we assume
that the generator ${\hat G}$ of the evolution operator ${\hat U}=\exp
(-i{\hat G})$ belongs to this class of Hamiltonians.\\

We first briefly summarize the main results of Refs. \cite{wil2,wil3}. Let
$\rho '=q'/p'$ be a rational number sufficiently close to $\rho =q/p$ and
such that $p'\gg p$. The $p'$ bands for $\rho '=q'/p'$ can then be grouped
into $p$ ``clusters" of adjacent bands, where each cluster is associated
in a natural way with a band $b$ for $\rho =q/p$. Namely, the energy or
quasienergy interval covered by the bands in the cluster is relatively
close to that covered by band $b$; in addition, the total Chern index of
the cluster is equal to $\sigma_b$,
\begin{equation}\label{sc}
\sigma (C_b)\equiv \sum_{b'=d(b)}^{d(b)+N_b-1}\sigma '_{b'}=\sigma_b \
,
\end{equation}
where $C_b$ denotes the cluster corresponding to band $b$, $d(b)$ is the
label of the lowest band in the cluster, $N_b$ is the number of bands in
the cluster, and $\sigma '_{b'}$ is a Chern index for $\rho '=q'/p'$. The
spectrum and eigenstates in cluster $C_b$ can be approximately calculated
from an effective Hamiltonian, $H_{\rm eff}$, obtained by properly
quantizing the band function $\omega_b({\bf w})$. The effective scaled
$\hbar$ in this quantization turns out to be
\begin{equation}\label{eh}
\rho_{\rm eff}=\frac{q_{\rm eff}}{p_{\rm eff}}
=\frac{pq'-p'q}{p'\sigma_b+q'\mu_b}\ ,
\end{equation}
where $\mu_b$ is the integer uniquely determined from the Diophantine
equation $p\sigma_b+q\mu_b=1$. It is easy to show that the numerator and
denominator in the last fraction in (\ref{eh}) are relatively prime
integers. Eq. (\ref{eh}) then implies a simple formula for the number of
bands in the cluster, $N_b=p_{\rm eff}$,
\begin{equation}\label{Nb}
N_b=p'\sigma_b+q'\mu_b \ .
\end{equation}

Now, the existence of an approximate effective Hamiltonian $H_{\rm eff}$
for a cluster means that the space of states in band $b$ approximately
coincides with the space of states in the $N_b$ bands of cluster $C_b$.
Provided $\rho '$ is sufficiently close to $\rho$, any state in the
cluster is well approximated by a linear combination of states that belong
only to band $b$. This fact is expressed concisely by the statement that
the projection operator for band $b$ is approximately equal to that for
cluster $C_b$: 
\begin{equation}\label{po}
\frac{1}{\vert {\rm BZ}\vert}\int_{\rm BZ}d{\bf w} \vert\Psi_{b,{\bf
w}}\rangle\langle \Psi_{b,{\bf w}}\vert \approx {\cal
N}\sum_{b'=d(b)}^{d(b)+N_b-1} \frac{1}{\vert {\rm BZ'}
\vert}\int_{\rm BZ'} d{\bf w'} \vert\Psi '_{b',{\bf w'}}\rangle 
\langle \Psi '_{b',{\bf w'}}\vert \ ,
\end{equation}
where all the primed quantities refer to $\rho '$ and ${\cal N}$ is a
normalization constant that remains to be determined. Let us assume, for
simplicity, that the phase-space variables ${\bf R'}=(u',\ v')$ for $\rho
'$ ($[{\hat u}',\ {\hat v}']=2\pi i\rho '$) are related to the variables
${\bf R}=(u,\ v)$ for $\rho$ by ${\bf R'}=\sqrt{\rho /\rho '}\, {\bf R}$. 
Using then the expressions in Eqs. (\ref{QES}) and (\ref{cs}) for both
$\rho$ and $\rho '$, we immediately obtain from Eq. (\ref{po}) that
\begin{equation}\label{BHDc}
P_b({\bf R})=\frac{1}{\vert {\rm BZ}\vert} \int_{\rm BZ}d{\bf
w}\vert\Psi_{b,{\bf w}}({\bf R})\vert ^2 \approx
\frac{1}{N_b}\sum_{b'=d(b)}^{d(b)+N_b-1}\frac{1}{\vert {\rm BZ'}\vert}
\int_{\rm BZ'}d{\bf w'}\vert\Psi '_{b',{\bf w'}}({\bf R'})\vert ^2 =
P'_{C_b}({\bf R'})\ ,
\end{equation}
where the constant ${\cal N}$ in Eq. (\ref{po}) has been determined as
${\cal N}=1/N_b$ from the normalization condition (\ref{nr}). Eq.
(\ref{BHDc}) shows that the BHD $P_b({\bf R})$ for band $b$ is
approximately equal to the generalized BHD $P'_{C_b}({\bf R'})$ for the
cluster $C_b$. In the limit $\rho '\rightarrow\rho$, the space of the
cluster becomes identical to that of band $b$ [the approximate equality in
Eq. (\ref{po}) is replaced by an equality], and $P'_{C_b}({\bf
R'})\rightarrow P_b({\bf R})$. This expresses the continuity of the
generalized BHD, associated with clusters $C_b$, under small variations of
$\rho$ on the rationals.\\

The most important reference values of $\rho$ are, of course, those with
$q=1$ (i.e., $\rho =1/p$), for which the topological characterization of
the classical-quantum correspondence on the torus is well established
\cite{le1,le2,da1}. Using the analysis in the previous paragraph, this
characterization can be easily extended to rational values $\rho '=q'/p'$
($q'\neq 1$) sufficiently close to $\rho =1/p$. Consider first the case of
$\sigma_b=0$, with eigenstates localized on a regular classical orbit and
exhibiting weak sensitivity to variations in the BCs. Here, the
Diophantine equation $p\sigma_b+q\mu_b=1$ ($q=1$) implies that $\mu_b=1$.
Then, from Eq. (\ref{Nb}), the corresponding cluster $C_b$ consists simply
of $N_b=q'$ bands with total Chern index $\sigma (C_b)=0$ [see Eq.
(\ref{sc})]. As shown in Sec. IV, one can find for such a cluster (or
generalized band) a new basis (\ref{ut}) whose effective Chern indices all
vanish. The cluster can then be viewed as exhibiting ``weak" sensitivity
to variations in the BCs, in analogy to the single band $b$. In addition,
because of Eq. (\ref{BHDc}), the generalized BHD for the cluster is
approximately equal to the BHD for band $b$, and is therefore localized on
the same classical regular orbit. In the case of $\sigma_b\neq 0$, the
number $N_b$ of bands in the cluster is generally different from $q'$ and
varies with $b$. However, the Chern index $\sigma (C_b)$ of the cluster is
still equal to $\sigma_b$ [Eq. (\ref{sc})], and, like the single band $b$,
the cluster exhibits ``strong" sensitivity to variations in the BCs, in
the sense that a new basis (\ref{ut}) whose effective Chern indices all
vanish does not exist (see Sec. IV). If the BHD for band $b$ is spread
over the classically chaotic region, the same will be true for the
generalized BHD of the cluster. The topological characterization of the
classical-quantum correspondence on the torus is thus generalized to $\rho
'$ sufficiently close to $\rho =1/p$ by replacing single bands $b$ with
the corresponding clusters $C_b$.\\

As an example, we consider the case of $\rho =1/5$ for the KH model with
$\gamma =0.3$. While for this value of $\gamma$ the chaotic region
occupies a significant portion of the phase space (see Fig. 1), the Chern
indices turn out to be the same as in the integrable limit (Harper model),
i.e., $\sigma_b=0$ for $b=1,\ 2,\ 4,\ 5$ (``regular-motion" bands) and
$\sigma_3=1$ (``chaotic-motion" band). Similarly, for three rational
approximants $\rho '$ of $\rho$, $\rho '=2/11,\ 4/21,\ 6/31$, the Chern
indices are the same as those in the Harper model \cite{qh} (however, the
quasienergy spectra and states are, of course, completely different from
those in the integrable limit). Using then formula (\ref{Nb}), we find,
for these values of $\rho '$, that $N_b=q'$ for $b=1,\ 2,\ 4,\ 5$ while
$N_b=q'+1$ for $b=3$. Fig. 8(a) shows the BHD $P_b({\bf R})$ for $b=1$ and
the corresponding generalized BHDs $P'_{C_b}({\bf R'})$, calculated using
Eq. (\ref{BHDc}), along the symmetry line $u=v$. Similar results for $b=3$
are shown in Fig. 8(b). The convergence of $P'_{C_b}({\bf R'})$ to
$P_b({\bf R})$ as $\rho '$ approaches $\rho$ is clear in both cases.\\

\begin{center}{\bf VII. CONCLUSIONS}\\ \end{center}

The framework of quantum dynamics on a toral phase space offers the
possibility of classifying the eigenstates as ``regular" or ``chaotic" 
according to their sensitivity to variations in the boundary conditions
(BCs), i.e., variations of the quasicoordinate ${\bf w}$ in the Brillouin
zone (BZ). A measure of this sensitivity for a given band $b$ is the Chern
index $\sigma_b$, which determines the quasiperiodicity conditions
(\ref{per1}) and (\ref{per2}) satisfied by the band eigenstates. Provided
$\rho$ (the scaled $\hbar$) assumes a rational value of the form $1/p$
($p$ being a sufficiently large integer), eigenstates well localized on
regular classical orbits belong to bands with $\sigma_b=0$. On the other
hand, eigenstates concentrated in chaotic regions of the phase space
usually belong to bands with $\sigma_b\neq 0$.\\

By definition, $\sigma_b$ characterizes the entire band $b$, rather than
individual eigenstates. The Chern-index classification into ``regular" or
``chaotic" refers then to bands and not to eigenstates. It is also
well known \cite{le1,da1} that $\sigma_b$ can be expressed as a uniform
average over ${\bf w}$ in the BZ:   
\begin{equation}\label{chb}
\sigma_b= \frac{i}{2\pi}\int_{\rm BZ} d{\bf w}\int_{T^2_Q}d^2{\bf R} 
\left [ \frac{\partial\Psi^{\ast}_{b,{\bf w}}({\bf R})}{\partial w_1}
\frac{\partial\Psi_{b,{\bf w}}({\bf R})}{\partial w_2} -
\frac{\partial\Psi^{\ast}_{b,{\bf w}}({\bf R})}{\partial w_2}
\frac{\partial\Psi_{b,{\bf w}}({\bf R})}{\partial w_1} \right ] \ ,
\end{equation} 
where $\Psi_{b,{\bf w}}({\bf R})$ is the coherent-state representation of
the eigenstate $\vert\Psi_{b,{\bf w}}\rangle$. These observations motivate
one to introduce other global characterizations of the bands which take
into account all the BCs only in an average sense, thus resolving the
ambiguity in the choice of a particular BC. These additional
characterizations should provide a more detailed description of the
classical-quantum correspondence on the torus without referring to the
individual, nonclassical BCs.\\

In this paper, we have introduced and studied one such characterization,
the band Husimi distribution (BHD), given by the uniform average of
$\vert\Psi_{b,{\bf w}}({\bf R})\vert ^2$ over ${\bf w}$. The BHD may be
viewed as the representative Husimi distribution for ``level" $b$ and is
expected to be closer to a classical probability density than
$\vert\Psi_{b,{\bf w}}({\bf R})\vert ^2$. This expectation was shown to be
fulfilled in several aspects in Sec. III. The BHD concept plays an
important role in extending the Chern-index characterization of the
classical-quantum correspondence to rational values of $\rho$ with
numerators larger than 1 (see Sec. VI). In this case, a generalized
version of the BHD has to be introduced (see Sec. IV), given by the
average of the BHDs associated with a cluster of adjacent bands for $\rho
'=q'/p'\approx 1/p$.\\

Smoothing a quantum probability distribution over a range of energy levels
is important in the theoretical study of scars using the semiclassical
periodic-orbit theory \cite{bog,ber}. In the case of the BHD, this
smoothing is performed over the continuous range of one band,
corresponding essentially to a single level in the framework of a toral
phase space. Using the adaptation of periodic-orbit theory to this
framework \cite{lem}, it may be possible to achieve a better understanding
of the nature of BHDs and generalized BHDs in the semiclassical limit.\\

An important question is to what extent information about the individual
band eigenstates can be recovered from the BHD. We recall here the well
known fact in solid-state physics that all the Bloch eigenstates in a band
can be simply reproduced from the corresponding Wannier function.  Thus,
in this case, no information about the individual eigenstates is lost by
averaging over them (the average being the Wannier function). Rel. 
(\ref{sas}), valid only for $\sigma_b=0$, is analogous to the formula
expressing the Bloch eigenstates in terms of the Wannier function. In this
case, Rel. (\ref{BHD0}) shows that the BHD is closely related to the
absolute value squared of a Wannier function. This relation may be
extended to $\sigma_b\neq 0$ using results from Ref. \cite{wil3}. It was
also shown recently \cite{dfw} that the Wigner functions of all of the
band eigenstates can be reproduced from the Wigner analog of the BHD. On
the basis of these observations, we expect that it should be possible to
recover, at least partially, relevant information about the individual
eigenstates from the BHDs. This will be investigated in future works.\\

\newpage 

{\bf Acknowledgments}\\
 
The authors would like to thank J. Zak, P. Leboeuf, M. Wilkinson, and
D. Arovas for useful comments and discussions. This work was
partially supported by the Israel Ministry of Science and Technology
and the Israel Science Foundation administered by the Israel Academy
of Sciences and Humanities.

\figure{FIG. 1. Classical Poincar\'{e} maps of typical orbits of the
kicked-Harper (KH) model (\ref{kh}) for different values of the
nonintegrability parameter $\gamma =A\tau /(2\pi I)$: (a) $\gamma =0.001$
(nearly-integrable regime). (b) $\gamma =0.26$ (mixed regime). Notice, for
future reference, the island chains of periods $6$ and $8$ surrounding the
central elliptic point. (c) $\gamma =0.56$. The chaotic region occupies a
large fraction of the phase space. (d) $\gamma =0.95$ (strongly-chaotic
regime).\label{f1}}

\figure{FIG. 2. Density plots of Husimi distributions $\vert\Psi_{b,{\bf
w}}({\bf R})\vert ^2$ for the KH model with parameters $\gamma =0.26$
[compare with Fig. 1(b)] and $\rho =1/11$. The central band $b=6$ (with
Chern index $\sigma_6=1$) is considered for different values of ${\bf w}$
in the BZ: (a) ${\bf w}=(0,\ 0)$, (b) ${\bf w}=(\pi /11,\ 0)$, (c) ${\bf
w}=(0,\ \pi /11)$, and (d) ${\bf w}=(\pi /11,\ \pi /11)$. In these plots,
as well as in Figs. 3-5, we use a power-law density scale (with power
$\approx 1/3$) and ten gray tones, with darker tones corresponding to
higher values of the distribution. In cases (b) and (c),
$\vert\Psi_{b,{\bf w}}({\bf R})\vert ^2$ is concentrated on one of the two
hyperbolic fixed points, while in cases (a) and (d) it is concentrated on
both points.\label{f2}}

\figure{FIG. 3. Similar to Fig. 2, but for band $b=5$ with Chern index
$\sigma_5=0$. Despite the fact that $\sigma_5=0$, we observe strong
sensitivity to variations in ${\bf w}$, as in the case of Fig. 2. In plot
(a) [${\bf w}=(0,\ 0)$], $\vert\Psi_{b,{\bf w}}({\bf R})\vert ^2$ is
concentrated on the island chain of period $8$ [see Fig. 1(b)]. Notice
that the localization regions in the case of plots (b) and (c) are
reversed relative to the corresponding plots in Fig. 2. Plot (d), on the
other hand, is qualitatively similar to Fig. 2(d).\label{f3}}

\figure{FIG. 4. Band Husimi distribution (BHD) for the case considered in
Fig. 2 ($b=6$). The BHD was calculated by averaging $\vert\Psi_{b,{\bf
w}}({\bf R})\vert ^2$ over $20\times 20$ values of ${\bf w}$, uniformly
distributed in the BZ. This BHD, with a localization region similar to
that of Figs. 2(a) and 2(d), is close to a classical probability
distribution for the stochastic layer in Fig. 1(b) (see text).\label{f4}}

\figure{FIG. 5. Similar to Fig. 4, but for the case considered in Fig. 3
($b=5$). The BHD appears to be concentrated on four of the six islands of
an island chain of period $6$, surrounding the main island chain of this
period in Fig. 1(b). The localization region for this BHD is
representative for band $b=5$ but is qualitatively different from that of
$\vert\Psi_{b,{\bf w}}({\bf R})\vert ^2$ for all the special values of
${\bf w}$ considered in Fig. 3.\label{f5}}

\figure{FIG. 6. (a) BHDs along the symmetry line $u=v$ for bands $b=3$
(solid line) and $b=4$ (dashed line) in the case of the KH model with
parameters $\gamma =0.26$ and $\rho =1/11$. The BHDs were calculated by
averaging $\vert\Psi_{b,{\bf w}}({\bf R})\vert ^2$ over $100\times 100$
values of ${\bf w}$ uniformly distributed in the BZ. (b) Similar to (a),
but for $\gamma =0.2645>\gamma_0\approx 0.264$, where $\gamma_0$ is a
degeneracy point for bands $b=3,\ 4$. Notice the dramatic change of the
BHDs relative to those of case (a), despite the small variation in
$\gamma$ through the degeneracy point. (c) Generalized BHD for bands
$b=3,\ 4$, obtained by averaging the results for the two bands in case (a)
(solid line) and in case (b) (dashed line).\label{f6}}

\figure{FIG. 7. (a) Similar to Fig. 6(a), but for the bands $b=5$ (solid
line), $b=6$ (dashed line), and $b=7$ (dotted line) in the case of $\gamma
=0.3385<\gamma_0\approx 0.3387$, where $\gamma_0$ is a degeneracy point of
all the three bands. (b) Similar to (a), but for $\gamma =0.339>\gamma_0$. 
(c) Generalized BHD for bands $b=5,\ 6,\ 7$, obtained by averaging the
results for the three bands in case (a) (solid line) and in case (b)
(dashed line, essentially indistinguishable from the solid
line).\label{f7}}

\figure{FIG. 8. BHDs along the symmetry line $u=v$ for bands and clusters
of bands in the KH model with nonintegrability parameter $\gamma =0.3$:
(a) The upper curve (see a magnification in the inset) is the BHD for band
$b=1$ in the case of $\rho =1/5$. The other three curves are the
generalized BHDs for the corresponding cluster $C_b$ in the cases of $\rho
=6/31,\ 4/21,\ 2/11$, in order of descending curves (see also inset). (b)
Similar to (a), but for the central band $b=3$. In both cases, the BHDs
were calculated by averaging $\vert\Psi_{b,{\bf w}}({\bf R})\vert ^2$ over
$400\times 400$ values of ${\bf w}$ uniformly distributed in the
BZ.\label{f8}}
 

\begin{references}
\bibitem{not} We have reconsidered the case of $\rho =1/11$ for the KH
model, first studied in Ref. \cite{le2}, by performing accurate
calculations of the quasienergy band structure and Chern indices for many
values of $\gamma$ in a sufficiently large interval. In particular, we
have found that at the degeneracy between bands $b=3,\ 4$ near
$\gamma=0.264$, the Chern indices change from $\sigma_3=\sigma_4=0$ to
$\sigma_3=-\sigma_4=2$, and not to $\sigma_3=-\sigma_4=1$, as reported in
Ref. \cite{le2}.
\end{references}
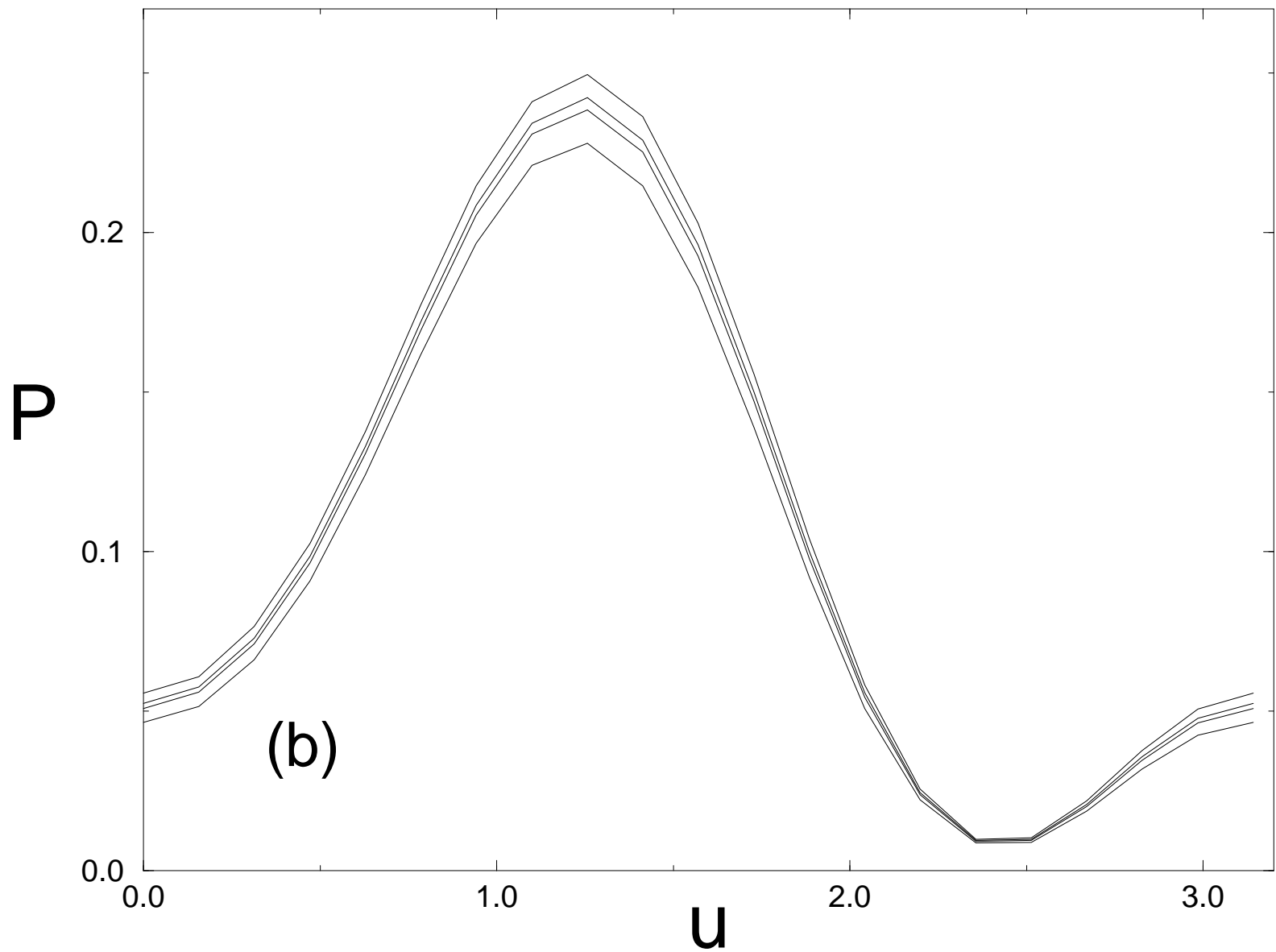
\end{document}